\documentclass[letterpaper]{article} % DO NOT CHANGE THIS
\usepackage[draft]{aaai25}  % DO NOT CHANGE THIS
\usepackage{times}  % DO NOT CHANGE THIS
\usepackage{helvet}  % DO NOT CHANGE THIS
\usepackage{courier}  % DO NOT CHANGE THIS
\usepackage[hyphens]{url}  % DO NOT CHANGE THIS
\usepackage{graphicx} % DO NOT CHANGE THIS
\usepackage{natbib}  % DO NOT CHANGE THIS AND DO NOT ADD ANY OPTIONS TO IT
\usepackage{caption} % DO NOT CHANGE THIS AND DO NOT ADD ANY OPTIONS TO IT
\usepackage{algorithm}
\usepackage{algorithmic}
\usepackage{amsmath}
\usepackage{amsthm}
\usepackage{float} %设置图片浮动位置的宏包
\usepackage{centernot}
\usepackage{multirow} %设置表格
\usepackage{array}
\usepackage{subfigure}
\usepackage{xcolor}
\usepackage{color}
\usepackage{float}
\newcommand{\CI}{\mathrel{\perp\mspace{-10mu}\perp}}

\newtheorem{theorem}{Theorem}

\usepackage{newfloat}
\usepackage{listings}
\DeclareCaptionStyle{ruled}{labelfont=normalfont,labelsep=colon,strut=off} % DO NOT CHANGE THIS
\floatstyle{ruled}
\newfloat{listing}{tb}{lst}{}
\floatname{listing}{Listing}
\title{Debiased Contrastive Representation Learning for Mitigating Dual Biases in Recommender Systems}
\author{
    Zhirong Huang\textsuperscript{\rm 1,\rm 2},
    Shichao Zhang\textsuperscript{\rm 1,\rm 2}\thanks{Corresponding author},
    Debo Cheng\textsuperscript{\rm 3}$^{*}$ ,
    Jiuyong Li\textsuperscript{\rm 3},
    Lin Liu\textsuperscript{\rm 3}, Guixian Zhang\textsuperscript{\rm 4}
}
\affiliations{
    \textsuperscript{\rm 1}1 Key Lab of Education Blockchain and Intelligent Technology, Ministry of Education, \\Guangxi Normal University, Guilin, 541004, China\\ 
\textsuperscript{\rm 2}Guangxi Key Lab of Multi-Source Information Mining and Security,\\ Guangxi Normal University, Guilin, 541004, China\\
    \textsuperscript{\rm 3}UniSA STEM, University of South Australia,\\ Mawson Lakes, Adelaide, Australia\\
    \textsuperscript{\rm 4}School of Computer Science and Technology, China University of Mining and Technology,\\ Xuzhou, Jiangsu, 221116, China\\
    zhangsc@mailbox.gxnu.edu.cn, chedy055@mymail.unisa.edu.au\\
}

\usepackage{bibentry}
\begin{document}

\maketitle

\begin{abstract}
 In recommender systems, popularity and conformity biases undermine recommender effectiveness by disproportionately favouring popular items, leading to their over-representation in recommendation lists and causing an unbalanced distribution of user-item historical data. We construct a causal graph to address both biases and describe the abstract data generation mechanism. Then, we use it as a guide to develop a novel Debiased Contrastive Learning framework for Mitigating Dual Biases, called DCLMDB. In DCLMDB, both popularity bias and conformity bias are handled in the model training process by contrastive learning to ensure that user choices and recommended items are not unduly influenced by conformity and popularity. Extensive experiments on two real-world datasets, Movielens-10M and Netflix, show that DCLMDB can effectively reduce the dual biases, as well as significantly enhance the accuracy and diversity of recommendations.
\end{abstract}


\section{Introduction}

Recommender systems are designed to predict user preferences and recommend items that might be of interest to users. They are widely used in e-commerce (e.g. by Amazon.com)~\cite{shoja2019customer}, streaming services (e.g. by Netflix)~\cite{gomez2015netflix}, social media platforms~\cite{liao2022sociallgn}, and other online services where personalised content is crucial~\cite{covington2016deep,xie2021explore}. In an era of information, recommender systems are increasingly indispensable since they not only assist users in finding content that aligns with their preferences but also offer substantial commercial benefits~\cite{shoja2019customer}. For example, e-commerce platforms can significantly increase the transaction probability by recommending products of interest to users. 

 \begin{figure}[t]
    \begin{center}
        \subfigure[]{
        \includegraphics[width=0.125\textwidth]{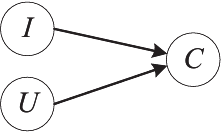}
        }
        \quad
        \quad
        \subfigure[]{
        \includegraphics[width=0.16\textwidth]{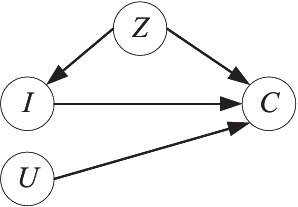}
        }
        \caption{Two causal graphs are used to show the recommendation process. $U$: user preference, $I$: exposed item, $C$: choice, $Z$: item popularity. (a) Traditional recommender methods; (b) Popularity bias caused by item popularity $Z$.}  
        \label{Fig1}
    \end{center}
\end{figure} 

Traditional recommender system models operate under the assumption that observational data is generated when a user's preference aligns with the attributes of an item. Various models employing collaborative filtering algorithms have been developed based on this premise~\cite{liang2018variational,zou2020neural,ji2020dual}. These models predict the likelihood of a user choosing an item by calculating the inner product of their respective embedding. We illustrate these models using the causal graph in Fig.~\ref{Fig1} (a), where $U$ and $I$ represent user preference and exposed item, respectively, and both are causes of $C$, the choice. However, the traditional modelling approach can be biased towards popular items. Popularity bias leads to over-recommendation of certain items to users despite users' lack of prior interaction with similar items, thereby missing the opportunity of recommending users truly interesting items by matching user preferences and item attributes.

Balanced representation of items in a recommender system is a common approach to addressing popularity bias~\cite{zhao2022popularity,schnabel2016recommendations}. For instance, Zhang et al.~\cite{zhang2021causal} proposed a novel training and inference paradigm called Popularity-bias Deconfounding and Adjusting (PDA). This method employs do-calculus~\cite{pearl2009causality,cheng2024data} to mitigate the negative effects of popularity bias during the training phase and adjusts predicted item popularity scores during inference for endowing recommendation policy with the desired level of popularity bias. PDA incorporates an item popularity node $Z$ into the traditional recommender model, as illustrated in Fig.~\ref{Fig1} (b). Specifically, $Z \rightarrow I$ represents that the popularity of an item influences its exposure rate, while $Z \rightarrow C$ indicates that item popularity impacts user choice because users tend to believe that a popular item has a high quality.

Conformity bias occurs when users align their choices with the group, even the item attributes conflict with their personal preferences~\cite{chen2023bias}. Zheng et al.~\cite{zheng2021disentangling} presented an innovative causal graph that outlines user-item interactions influenced by conformity. They developed a framework called Disentangling Interest and Conformity with Causal Embedding (DICE), which utilises different embedded representations to independently capture user interest and conformity, thereby reducing the impact of user conformity bias. 
 
\begin{figure}[t]
    \begin{center}
    \subfigure[]{
        \includegraphics[width=0.16\textwidth]{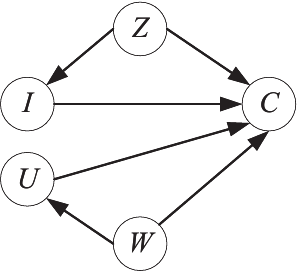}
        }
        \quad
        \quad
        \subfigure[]{
        \includegraphics[width=0.16\textwidth]{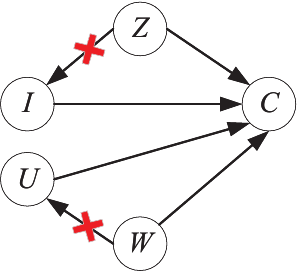}
        }
    \caption{Causal graphs showing both popularity bias and conformity bias and an illustration of our solution. $U$: user preference, $I$: exposed item, $C$: choice, $Z$: item popularity, $W$: conformity influence. (a) The causal graph considers the effects of popularity items and conformity influence; (b) we cut off the edges $Z\rightarrow I$ and $W\rightarrow U$ in the training model.}
    \label{Fig2} 
    \end{center}
\end{figure}

Existing methods address either popularity bias or conformity bias, but no works deal with both biases. Popularity and conformity biases often coexist (as verified through experiments detailed in the Appendix) and should be addressed simultaneously. In this paper, we aim to tackle the complexity of user-item interactions to mitigate both biases. We model both biases using the causal graph as shown in Fig.~\ref{Fig2} (a). Specifically, $W \rightarrow U$ demonstrates that conformity distorts user judgement, while $W \rightarrow C$ denote that conformity directly impacts user choice, e.g., word-of-mouth. %Therefore, a comprehensive approach addressing these dual biases from both item and user viewpoints is crucial for developing effective personalised recommendation systems.

To tackle the issue of the dual biases, it is essential to sever connections that contribute to these biases. That is, we need to cut the edges $Z\rightarrow I$ and $W\rightarrow U$ as shown in the manipulated causal graph~\cite{pearl2009causality} in Fig.~\ref{Fig2} (b) when training a recommender model. In this way, user choices and recommended items are not influenced by conformity and popularity. To achieve this, we propose a novel framework, the Debiased Contrastive Learning framework for Mitigating Dual Biases (DCLMDB) to learn and disentangle the latent representations $Z$ and $W$ from user-item interactions, aiming to remove the connections contributing to the biases. Fine-tuning $Z$ and $W$ to address the dual biases by applying back-door adjustment~\cite{pearl2009causality} improves the implementation of the click prediction task and enhances recommendation accuracy.
% By applying the back-door adjustment~\cite{pearl2009causality}, both representations $Z$ and $W$ are fine-tuned to address the dual biases, thereby enhancing the accuracy of the recommendations.
The main contributions of our work are summarised as follows:
\begin{itemize}
    \item[\textbullet] We study the problem regarding popularity and conformity biases in recommender systems and use a graphical causal modelling approach to address the problem. To the best of our knowledge, this is the first work which simultaneously addresses both the popularity bias associated with items and the conformity bias stemming from users.
    \item[\textbullet] We design and develop a novel debiased contrastive learning framework, DCLMDB, for mitigating both popularity and conformity biases by learning two embeddings derived from the latent space of items and users.
    \item[\textbullet] Extensive experiments conducted on two real-world datasets validate the effectiveness and robustness of our DCLMDB model.
\end{itemize}

\section{Related Work}

Recommender systems primarily aim to predict user choices. Collaborative filtering, which leverages user-item historical data to uncover user-item similarities, remains a dominant approach for personalised recommendations. In recent years, causal-based recommendation methods have emerged. This section reviews related work in two main areas: traditional and causal recommendation methods.

\subsection{Traditional Recommendation Methods}
Traditional collaborative filtering methods focus on learning user and item embeddings to make predictions. Early methods, such as Matrix Factorisation (MF)~\cite{koren2009matrix}, decompose the user-item rating matrix to predict user ratings and personalised rankings. However, MF is not inherently optimised for personalised ranking, leading to the development of the Bayesian Personalised Ranking (BPR) loss by Rendle et al.~\cite{rendle2012bpr}, which has become a standard in personalised recommender methods. With the advances of deep learning, He et al.~\cite{he2017neural} proposed the Neural network-based Collaborative Filtering (NCF) framework, replacing the inner product with a multi-layer perceptron to model user-item choices. 

To enhance the capture of interaction information between users and items, researchers have incorporated graph structures into recommender systems~\cite{xia2022hypergraph,zhu2021graph,liu2021interest}. Wang et al.~\cite{wang2019neural} introduced the Neural Graph Collaborative Filtering (NGCF) framework, grounded in Graph Convolutional Networks (GCN), which markedly enhances recommendation performance by more precisely embedding user-item interaction data. Nonetheless, He et al.~\cite{he2020lightgcn} noticed that the feature transformation and the nonlinear activation components in NGCF did not significantly improve the performance. Therefore, they retained only the neighbourhood aggregation component and
then proposed the Light Graph Convolutional Network (LightGCN). However, these methods often overlook popularity bias and conformity bias, inadvertently amplifying the biases during training and skewing recommendations towards popular items.
 
\subsection{Causal Recommendation Methods}
The impact of popularity bias and conformity bias has led to the emergence of causal inference-based solutions. A notable method is Inverse Propensity Scoring (IPS)~\cite{schnabel2016recommendations}, which reweights items based on their popularity, giving less popular items a greater weight to mitigate bias. Despite IPS's effectiveness, its high variance led to the development of variants for stability~\cite{bottou2013counterfactual,gruson2019offline,zhu2020unbiased}. CausE, proposed by Bonner et al.~\cite{bonner2018causal}, uses both a biased and a small unbiased dataset to obtain two sets of embeddings, which are later regularised to reduce their disparity. 

However, constructing unbiased datasets is costly and often ignores user conformity. Zheng et al.~\cite{zheng2021disentangling} tackled conformity bias at the embedding level with the framework that DICE, separating user and item embeddings into interest and conformity parts of the user by constructing specific training samples. Zhao et al.~\cite{zhao2023disentangled} argued that the DICE approach to constructing specific training samples introduces noise, so they modified it to use contrast learning for data augmentation and then decoupled user interest and conformity based on DICE. On the other hand, Zhang et al.~\cite{zhang2021causal} proposed the PDA method, which uses the do-calculus to eliminate the popularity bias on the item side during the model training phase. 
% Zhao et al.~\cite{zhao2022popularity} argued that PDA might not effectively ensure the injection of benign popularity bias and proposed a novel framework termed Time-aware DisEntangled (TIDE), utilizing time information to differentiate between benign and harmful popularity biases in items. 
These approaches, while innovative in addressing item-based popularity bias or conformity of user, have not considered mitigating both biases simultaneously. 

\section{The Proposed DCLMDB Framework}
We first provide the problem definition and then analyse the effects of both popularity bias and conformity bias on the effectiveness of recommendations from a causal perspective. Subsequently, we propose DCLMDB, the debiasing framework based on disentangled contrastive learning designed to simultaneously mitigate the negative impacts of popularity bias and conformity bias in recommender systems. We provide definitions/concepts of causality related to our DCLMDB framework in the Appendix due to the page limit.

% we propose a novel Debiased Contrastive Learning framework for Mitigating Dual Biases (DCLMDB), 

\begin{figure*}[htbp]
    \centering
    \includegraphics[width=0.7\textwidth]{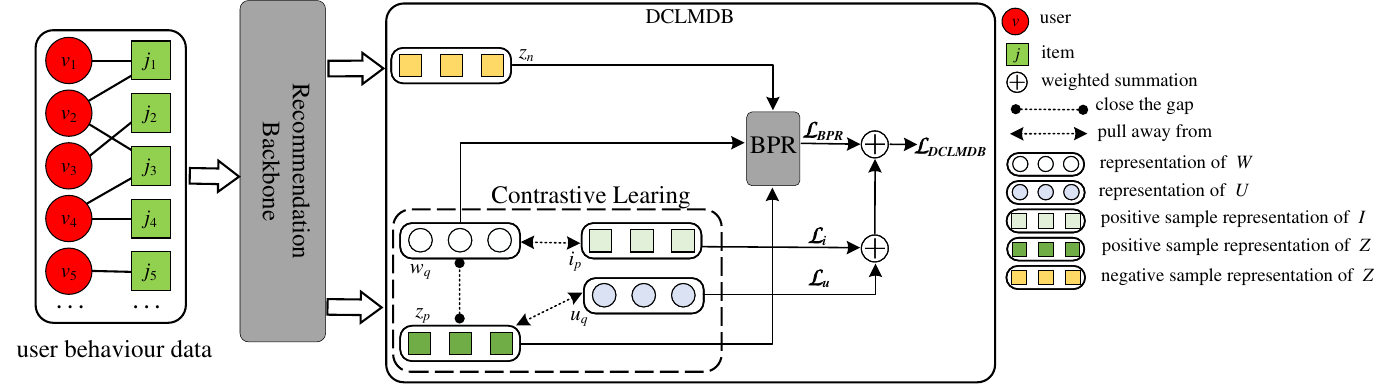}
    \caption{Overall structure of the proposed DCLMDB. First, we use the backbone to represent the input data as base embeddings ($U$ and $I$) and debiased embeddings ($Z$ and $W$). Both sets of embeddings encompass those of users and items. Subsequently, in the debiasing learning phase (i.e., the dotted box portion of the figure), we employ contrastive learning to steer the debiased embeddings away from biases inherent in the base embeddings. The specific realisations of $\mathcal{L}_{u}$, $\mathcal{L}_{i}$ and $\mathcal{L}_{BPR}$ are in Eq. (\ref{eq008}), Eq. (\ref{eq009}) and Eq. (\ref{eq0010}) respectively. Finally, $\mathcal{L}_{u}$, $\mathcal{L}_{i}$ and $\mathcal{L}_{BPR}$ are summed as in Eq. (\ref{eq0011}) to obtain the final loss function $\mathcal{L}_{DCLMDB}$.}
    \label{Fig4}
\end{figure*}

\subsection{Problem Setting and A Causal Analysis}
In a recommender system, there are two main sets: $V$, the set of users and $J$, the set of items. We denote a specific user in set $V$ as $v$. Within set $J$, we identify two types of items, $p$ and $n$, where $p$ is an item that user $q$ has chosen, while $n$ are many items that the user $q$ did not choose. Let $D$ denote user behaviour data, which can be represented as a set of triples, i.e., $D=\{(v, p, n) | p,n \in J, v \in V\}$. We use $C$ to indicate whether or not a user has chosen an item. Due to the complexity of user-item interplay, the historical data $D$ often does not reflect real user preferences and exposed items. Thus, we need to reconstruct both user preference (denoted as $U$) and exposed items (denoted as $I$) from $D$.

In this work, we simultaneously consider both item popularity and user conformity biases. To analyse the dual biases, we propose a new causal graph $\mathcal{G}$, depicted in Fig.~\ref{Fig2} (a), which explicates the factors contributing to popularity bias and conformity bias in recommender systems. In the causal graph $\mathcal{G}$, we use $Z$ to represent an item's popularity, which is considered as a latent factor. Traditional methods, such as matrix factorisation, do not explicitly model this aspect, yet it significantly impacts the effectiveness of the recommendation. $W$ denotes conformity influence, a latent factor reflecting the behaviour of other users who choose the item. The relationships between nodes in the causal graph $\mathcal{G}$ are represented by edges, which are explained as follows:

\begin{itemize}
\item[\textbullet] $\left(U, I, Z, W\right)\rightarrow C$ denotes that the four edges pointing toward $C$ from $U$, $I$, $Z$ and $W$ respectively, i.e., $C$ is determined by the four factors: $U$, $I$, $Z$ and $W$. Traditional recommendation methods operate under the assumption that a choice $C$ occurs when user preference $U$ matches with attributes of the exposed item $I$. In this work, we aim to learn the two latent causes of $C$, $Z$ and $W$, to account for popularity and conformity biases. Specifically, the edge $Z \rightarrow C$ signifies that an item's popularity influences the choice. For example, a ``popular movie'' is more likely to be chosen by a user. Similarly, $W \rightarrow C$ refers to the effect of user word-of-mouth on choice, e.g., a movie with high positive ratings is more likely to be chosen by a user.
\item[\textbullet] $W \rightarrow U$ represents the influence of conformity on a user. This conformity effect can be detrimental to the recommender system as it may not accurately reflect the user's genuine preference. For instance, a user might conform to many other users on an online forum and choose the same movie, even though it is not their preferred type of movie.
\item[\textbullet] $Z \rightarrow I$ indicates that the popularity of items affects their exposure. For example, on many online movie ticket websites, merchants can have their movies appear on the homepage recommendation list by purchasing certain traffic services from the website (i.e., increase item exposure). However, this does not reflect the attributes and quality of the items; it is merely a means to increase the sales of movies. 
\end{itemize}

From the causal graph $\mathcal{G}$, we know that $Z$ and $W$ are confounders between $\left(I, C\right)$ and $\left(U, C\right)$, respectively. Thus, $W$ affects the observed choices through the causal paths $W \rightarrow C$ and $W \rightarrow U \rightarrow C$. The path $W \rightarrow U \rightarrow C$ indicates that the conformity effect contributes to a higher prevalence of certain choices, leading to what is known as conformity bias amplification~\cite{gomez2015netflix,chen2023bias}. This effect is undesirable in a recommender system. Similarly, $Z$ affects the observed choices through $Z \rightarrow C$ and $Z \rightarrow I \rightarrow C$. The path $Z \rightarrow I \rightarrow C$ indicates that the popularity of an item increases its exposure, leading to a higher prevalence of popular items in observed choices. This phenomenon results in popularity bias amplification, which is undesirable in a recommender system~\cite{zhang2021causal}. An effective recommender system should accurately estimate a user's preferences and recommend the appropriate quality items. However, popularity and conformity biases lead to a false reflection of item quality or the user's genuine preference. Hence, they need to be mitigated to enhance a recommender system's effectiveness.

\subsection{Mitigating the Dual Biases with Causal Inference}
Guided by the proposed causal graph $\mathcal{G}$ in Fig.~\ref{Fig2} (a) and the causal analysis in the previous section, we aim to design a data-driven method to mitigate both the popularity and conformity biases in recommender systems. As indicated in Fig.~\ref{Fig2} (b), to deal with the biases, we need to remove the influence of popularity ($Z$) and conformity ($W$) on item exposure ($I$) and user preferences ($U$), respectively. To this end, we propose to perform $do (I, U)$, i.e., the do-operation on $I$ and $U$~\cite{pearl2009causality}. The ``$do$'' operator, denoted as $do(X = x)$, or $do(x)$ for short, represents an intervention where $X$ is set to a specific value $x$ intentionally, rather than by observing $X$ naturally occurring at $x$. 
% This operation separates mere correlations from actual causal effects~\cite{pearl2009causality,cheng2024data}. 

Applying the do operation on $I$ and $U$ removes all the edges pointing to $I$ and $U$, i.e., the edges $Z \rightarrow I$ and $W \rightarrow U$ from $\mathcal{G}$ as shown in Fig.~\ref{Fig2} (b) by the red crosses. We denote the manipulated graph as $\mathcal{G}_{\overline{U, I}}$. Note that cutting off the edges $Z \rightarrow I$ and $W \rightarrow U$ from $\mathcal{G}$ to obtain $\mathcal{G}_{\overline{U, I}}$ is equivalent to obtaining $P(C\mid do(U, I))$ from data. Thus, as implied by the manipulated graph $G_{\bar{U, I}}$, we need to ensure that $W$ and $U$ are independent, as well as $Z$ and $I$ when we learn the four embeddings from $D$. In the following sections, we will introduce the details of our method for achieving debiased recommendation based on causal manipulation.

\subsection{Debiased Contrastive Representation Learning}
\label{subsec:DeCRL}
Based on the above theoretical analysis from the perspective of causal inference, we proposed the novel Debiased Contrastive Learning framework for Mitigating Dual Biases (DCLMDB) as outlined in Fig.~\ref{Fig4}. 

It is challenging to learn both latent embeddings $Z$ and $W$ at the same time because, as shown in the causal graph in Fig.~\ref{Fig2} (b), both are latent factors encoded within the user-item historical data and are influenced by the interactive nature of user-item information. To tackle this challenge, we employ contrastive representation learning~\cite{schroff2015facenet} to learn the embeddings $Z$ and $W$ derived from the latent spaces of items and users, guided by the proposed causal graph in Fig.~\ref{Fig2} (b) to effectively mitigate the dual biases in recommender systems. Contrastive learning reduces the correlation between positive and negative samples in the feature space by adjusting the model such that the distance between the anchor point and positive samples (which are drawn closer together) is smaller than the distance between the anchor point and negative samples (which are pushed further apart).

First, we utilise a normal distribution to initialise the four embeddings: $Z$, $W$, $U$, and $I$. Then, we impose constraints in our DCLMDB method by ensuring that the learned $Z$ is independent of $I$ and $W$ is independent of $U$. To achieve these independencies, contrastive representation learning is employed to distance $Z$ (or $W$) from $I$ (or $U$), thereby keeping $Z$ (or $W$) away from the biases in $I$ (or $U$) during the training phase. In our DCLMDB, $I$ and $U$ are used as negative samples for training $Z$ and $W$ through contrastive learning. We define the following similarity functions to calculate the similarity between the pairs $(W, Z)$, $(U, Z)$, and $(W, I)$:
	\begin{equation}
        \label{eq005}
		\begin{array}{l}
			\mathcal{S}_{wz}=\left<w_v, z_p\right>,
		\end{array}
	\end{equation}	
	\begin{equation}
        \label{eq006}
		\begin{array}{l}
			\mathcal{S}_{uz}=\left<u_v, z_p\right>,
		\end{array}
	\end{equation}
	\begin{equation}
        \label{eq007}
		\begin{array}{l}
			\mathcal{S}_{wi}=\left<w_v, i_p\right>,
		\end{array}
	\end{equation}
\noindent where $\left<\cdot,\cdot\right>$ denotes the dot product operation and is used to measure matching scores among the sample pairs $(w_v, z_p)$, $(u_v, z_p)$, and $(w_v, i_p)$. $u_v$ and $w_v$ represent the embeddings of user $v$, while $i_p$ and $z_p$ are the embeddings of the positive sample of item $p$ (i.e., the user $v$ selected item).  We use two triplet loss functions in contrastive learning to maximise $S_{wz}$ and minimise $S_{wz}$ and $S_{wi}$, thereby reducing the distance between the anchor point and the positive samples while increasing the distance between the anchor point and the negative samples. 
Note that in the two loss functions, $w_v$ and $z_p$ serve as anchor points and positive samples, and $u_v$ and $i_p$ serve as negative samples. The two loss functions are defined as follows. 
    \begin{equation}
        \label{eq008}
		\begin{array}{l}
			\mathcal{L}_{u}=max\left(S_{wz}-S_{uz}+m,0\right),
		\end{array}
	\end{equation}
	\begin{equation}
        \label{eq009}
		\begin{array}{l}
			\mathcal{L}_{i}=max\left(S_{wz}-S_{wi}+m,0\right),
		\end{array}
    \end{equation}
\noindent where $m$ is a hyperparameter. By maximising these distances, we keep $w_v$ (or $z_p$) away from $u_v$ (or $i_p$), and thereby reducing biases in $u_v$ (or $i_p$). Consequently, DCLMDB yields two embeddings, $Z$ and $W$, such that is as uncorrelated with $I$ as possible, and $W$ is as uncorrelated with $U$ as possible. Our DCLMDB seeks to redirect $W$ and $Z$ away from the biases found in $U$ and $I$. Both loss functions $\mathcal{L}_{u}$ and $\mathcal{L}_{i}$ are used as regularisation terms in the click prediction task of the recommender system to ensure that $Z$ (or $W$) and $I$ (or $U$) remain independent during the training phase.
	
However, merely distancing $\left(W, Z\right)$ from $\left(U, I\right)$ does not guarantee the elimination of the dual biases. Since we are distancing them in multiple dimensions, it is not certain whether the direction of their distancing is mitigating or exacerbating the bias.
We use the Bayesian Personalised Ranking (BPR) loss function to guide the direction of the optimisation of the two embeddings $W$ and $Z$, as well as to achieve the main task (click prediction) in the recommender system.
	
	\begin{equation}
        \label{eq0010}
		\begin{array}{l}
			\mathcal{L}_{BRP}=-\displaystyle\sum_{\left(v,p,n\right)\in D}\ln\sigma\left(\left<w_v, z_p\right>-\left<w_v, z_n\right>\right),
		\end{array}
	\end{equation}

\noindent where $\sigma(\cdot)$ is the activation function, and $n$ denotes the negative sample item (i.e., the user $q$ unselected items). Each training instance within the BPR loss function is represented by a ternary $\left(v, p, n\right)$. This BPR function acts as the primary loss function in the recommender system, aiding in the click prediction. We utilise the Popularity-based Negative Sampling with Margin (PNSM) strategy~\cite{zheng2021disentangling} for selecting negative sample items. Consequently, the overall loss function of DCLMDB can be articulated as follows: 		
    \begin{equation}
        \label{eq0011}
		\begin{array}{l}
			\mathcal{L}_{DCLMDB}=\alpha\cdot \mathcal{L}_{BPR}+\beta \cdot \left(\mathcal{L}_u+\mathcal{L}_i\right),
		\end{array}
    \end{equation} 
\noindent where $\alpha$ and $\beta$ stand as hyperparameters. Note that $\mathcal{L}_{DCLMDB}$ is a realistic implementation of the causal graph in Fig.~\ref{Fig2} (b) that takes into account the reality of the click prediction task. For example, to leverage similar user behaviour for recommendations, we use the inner product operation instead of the distance metric in the latent space.  

To summarise, we use the embeddings $Z$ and $W$ to separate biased input data, combined with the corresponding user $v$ (i.e., $U$) and item $p$ (or $n$) (i.e., $I$) for click prediction to conform to the causal graph assumptions as shown in Fig.~\ref{Fig2} (b).
Additionally, $\mathcal{L}_{u}$ (or $\mathcal{L}_{i}$) serves as a regular term to ensure that $W$ (or $Z$) and $U$ (or $I$) remain independent. 
 
By combining the two contrastive learning loss functions with the BPR function, we obtain our ultimate loss function, $\mathcal{L}_{DCLMDB}$. This combination enables us to learn two embeddings, $Z$ and $W$, such that $Z$, $I$, $W$ and $U$ are mutually independent, as shown in the causal graph in Fig.~\ref{Fig2} (b). These independences align with the core objective of estimating click probability. Such alignment strengthens the direction of debiasing representation, ensuring that our DCLMDB effectively reduces bias without inadvertently intensifying it. Furthermore, our DCLMDB, being independent of specific data and models, serves as a generalised framework that can be seamlessly incorporated into various mainstream recommendation models.

\section{Experiments}
In this section, we conduct experiments on two real-world datasets to evaluate the performance of DCLMDB against state-of-the-art recommendation methods. The details of the parameter settings for all methods and additional experiments are provided in the Appendix due to the page limit.

\subsection{Experimental Settings}
\begin{table}[t]
		\centering
		\begin{tabular}{cccc} \hline
			Dataset & User & Item & Interaction  \\ \hline
			Movielens-10M & 37,962 & 4,819 & 1,371,473 \\ 
			Netflix & 32,450 & 8,432 & 2,212,690 \\ \hline
		\end{tabular}
        \caption{\centering Details of two real-world datasets}
        \label{Table 1}
    \end{table} 
\begin{table*}[!h]
		\centering
            \begin{tabular}{cc|cccccccc}
			\hline
			\multicolumn{2}{c|}{\multirow{2}{*}{Dataset}} &
			\multicolumn{8}{c}{Movielens-10M} \\ \cline{3-10}
			\multicolumn{2}{c|}{} &
			\multicolumn{4}{c|}{Top-K=20} &
			\multicolumn{4}{c}{Top-K=50} \\ \hline
			\multicolumn{1}{c|}{Backbone} &
			Method &
			\multicolumn{1}{c}{Recall$\uparrow$} &
			\multicolumn{1}{c}{HR$\uparrow$} &
			\multicolumn{1}{c}{NDCG$\uparrow$} &
			\multicolumn{1}{c|}{Imp.$\uparrow$} &
			\multicolumn{1}{c}{Recall$\uparrow$} &
			\multicolumn{1}{c}{HR$\uparrow$} &
			\multicolumn{1}{c}{NDCG}$\uparrow$ &
			Imp.$\uparrow$ \\ \hline
			\multicolumn{1}{c|}{\multirow{8}{*}{MF}} &
			Original &
			\multicolumn{1}{c}{0.1276} &
			\multicolumn{1}{c}{0.4397} &
			\multicolumn{1}{c}{0.0832} &
			\multicolumn{1}{c|}{--} &
			\multicolumn{1}{c}{0.2332} &
			\multicolumn{1}{c}{0.6308} &
			\multicolumn{1}{c}{0.1156} &
			-- \\ %\cline{2-10} 
			\multicolumn{1}{c|}{} &
			IPS &
			\multicolumn{1}{c}{0.1228} &
			\multicolumn{1}{c}{0.4210} &
			\multicolumn{1}{c}{0.0779} &
			\multicolumn{1}{c|}{-3.76\%} &
			\multicolumn{1}{c}{0.2168} &
			\multicolumn{1}{c}{0.6016} &
			\multicolumn{1}{c}{0.1070} &
			-7.03\%\\ %\cline{2-10} 
			\multicolumn{1}{c|}{} &
			IPS-C &
			\multicolumn{1}{c}{0.1277} &
			\multicolumn{1}{c}{0.4335} &
			\multicolumn{1}{c}{0.0809} &
			\multicolumn{1}{c|}{+0.08\%} &
			\multicolumn{1}{c}{0.2224} &
			\multicolumn{1}{c}{0.6150} &
			\multicolumn{1}{c}{0.1102} &
			-4.63\%\\ %\cline{2-10} 
			\multicolumn{1}{c|}{} &
			IPS-CN &
			\multicolumn{1}{c}{0.1494} &
			\multicolumn{1}{c}{0.4881} &
			\multicolumn{1}{c}{0.0991} &
			\multicolumn{1}{c|}{+17.08\%} &
			\multicolumn{1}{c}{0.2643} &
			\multicolumn{1}{c}{0.6640} &
			\multicolumn{1}{c}{0.1347} &
			+13.33\% \\ %\cline{2-10} 
			\multicolumn{1}{c|}{} &
			CausE &
			\multicolumn{1}{c}{0.1164} &
			\multicolumn{1}{c}{0.4144} &
			\multicolumn{1}{c}{0.0770} &
			\multicolumn{1}{c|}{-8.77\%} &
			\multicolumn{1}{c}{0.2076} &
			\multicolumn{1}{c}{0.5940} &
			\multicolumn{1}{c}{0.1047} &
			-10.98\% \\ %\cline{2-10} 
			\multicolumn{1}{c|}{} &
			DICE &
			\multicolumn{1}{c}{\underline{0.1626}} &
			\multicolumn{1}{c}{\underline{0.5202}} &
			\multicolumn{1}{c}{\underline{0.1076}} &
			\multicolumn{1}{c|}{\underline{+27.42\%}} &
			\multicolumn{1}{c}{\underline{0.2854}} &
			\multicolumn{1}{c}{\underline{0.6941}} &
			\multicolumn{1}{c}{\underline{0.1459}} &
			\underline{+22.38\%} \\ %\cline{2-10} 
            \multicolumn{1}{c|}{} &
			DCCL &
			\multicolumn{1}{c}{0.1503} &
			\multicolumn{1}{c}{0.4874} &
			\multicolumn{1}{c}{0.0975} &
			\multicolumn{1}{c|}{+17.79\%} &
			\multicolumn{1}{c}{0.2636} &
			\multicolumn{1}{c}{0.6676} &
			\multicolumn{1}{c}{0.1326} &
			+13.04\% \\ %\cline{2-10}
			\multicolumn{1}{c|}{} &
			 DCLMDB &
			\multicolumn{1}{c}{\pmb{0.1724}} &
			\multicolumn{1}{c}{\pmb{0.5415}} &
			\multicolumn{1}{c}{\pmb{0.1157}} &
			\multicolumn{1}{c|}{\pmb{+35.11\%}} &
			\multicolumn{1}{c}{\pmb{0.2948}} &
			\multicolumn{1}{c}{\pmb{0.7084}} &
			\multicolumn{1}{c}{\pmb{0.1539}} &
			\pmb{+26.42\%} \\ \hline
			\multicolumn{1}{c|}{\multirow{8}{*}{LightGCN}} &
			Original &
			\multicolumn{1}{c}{0.1462} &
			\multicolumn{1}{c}{0.4831} &
			\multicolumn{1}{c}{0.0952} &
			\multicolumn{1}{c|}{--} &
			\multicolumn{1}{c}{0.2631} &
			\multicolumn{1}{c}{0.6688} &
			\multicolumn{1}{c}{0.1316} &
			-- \\ %\cline{2-10} 
			\multicolumn{1}{c|}{} &
			IPS &
			\multicolumn{1}{c}{0.1298} &
			\multicolumn{1}{c}{0.4438} &
			\multicolumn{1}{c}{0.0849} &
			\multicolumn{1}{c|}{-11.22\%} &
			\multicolumn{1}{c}{0.2325} &
			\multicolumn{1}{c}{0.6196} &
			\multicolumn{1}{c}{0.1170} &
			-11.63\% \\ %\cline{2-10} 
			\multicolumn{1}{c|}{} &
			IPS-C &
			\multicolumn{1}{c}{0.1327} &
			\multicolumn{1}{c}{0.4533} &
			\multicolumn{1}{c}{0.0871} &
			\multicolumn{1}{c|}{-9.23\%} &
			\multicolumn{1}{c}{0.2383} &
			\multicolumn{1}{c}{0.6302} &
			\multicolumn{1}{c}{0.1201} &
			-9.43\% \\ %\cline{2-10} 
			\multicolumn{1}{c|}{} &
			IPS-CN &
			\multicolumn{1}{c}{0.1203} &
			\multicolumn{1}{c}{0.4220} &
			\multicolumn{1}{c}{0.0766} &
			\multicolumn{1}{c|}{-17.71\%} &
			\multicolumn{1}{c}{0.2396} &
			\multicolumn{1}{c}{0.6277} &
			\multicolumn{1}{c}{0.1132} &
			-8.93\% \\ %\cline{2-10}  
			\multicolumn{1}{c|}{} &
			CausE &
			\multicolumn{1}{c}{0.1164} &
			\multicolumn{1}{c}{0.4099} &
			\multicolumn{1}{c}{0.0727} &
			\multicolumn{1}{c|}{-20.38\%} &
			\multicolumn{1}{c}{0.2204} &
			\multicolumn{1}{c}{0.6080} &
			\multicolumn{1}{c}{0.1046} &
			-16.23\% \\ %\cline{2-10} 
			\multicolumn{1}{c|}{} &
			DICE &
			\multicolumn{1}{c}{\underline{0.1810}} &
			\multicolumn{1}{c}{\underline{0.5564}} &
			\multicolumn{1}{c}{\underline{0.1228}} &
			\multicolumn{1}{c|}{\underline{+23.80\%}} &
			\multicolumn{1}{c}{\underline{0.3109}} &
			\multicolumn{1}{c}{\underline{0.7219}} &
			\multicolumn{1}{c}{\underline{0.1632}} &
			\underline{+18.17\%}\\ %\cline{2-10} 
            \multicolumn{1}{c|}{} &
			DCCL &
			\multicolumn{1}{c}{0.1462} &
			\multicolumn{1}{c}{0.4824} &
			\multicolumn{1}{c}{0.0947} &
			\multicolumn{1}{c|}{{0\%}} &
			\multicolumn{1}{c}{0.2644} &
			\multicolumn{1}{c}{0.6711} &
			\multicolumn{1}{c}{0.1311} &
			+0.49\%\\ %\cline{2-10}
			\multicolumn{1}{c|}{} &
			 DCLMDB &
			\multicolumn{1}{c}{\pmb{0.1832}} &
			\multicolumn{1}{c}{\pmb{0.5601}} &
			\multicolumn{1}{c}{\pmb{0.1240}} &
			\multicolumn{1}{c|}{\pmb{+24.69\%}} &
			\multicolumn{1}{c}{\pmb{0.3110}} &
			\multicolumn{1}{c}{\pmb{0.7239}} &
			\multicolumn{1}{c}{\pmb{0.1637}} &
			\pmb{+18.21\%}\\ \hline
		\end{tabular}
        \caption{The performance of all methods on Movielens-10M. The ``Original" indicates that only the backbone is used, with no additional causal debiasing methods. The best results are highlighted in bold, and the second-best result is underlined. The DCLMDB's enhancement on the Movielens-10M is highly significant, exhibiting a $p$-value of 0.004 when pairwise compared to the 12 results of the second-best method. }
		\label{Table 2}	
     \end{table*}
\textbf{Datasets}: We utilised two real-world datasets: the Movielens-10M dataset~\cite{harper2015movielens} and the Netflix dataset~\cite{bennett2007netflix}. Both datasets comprise movie ratings featuring user IDs, movie IDs, and the ratings assigned by users. We preprocess the two datasets according to the previous methods~\cite{zheng2021disentangling}. We binarised both datasets by keeping the rating of five stars as one and others as zero. We randomly selected 40\% of the items (which can be seen as the outcome of a completely random recommendation strategy) and designated 10\% and 20\% of these as the validation and test sets, respectively. Finally, the remaining 10\% of the random data and another 60\% of the unselected samples form the training set. This approach to dataset handling aims for realism, ensuring that test data are not influenced by item popularity and that each item has an equal chance of being exposed to users. Table~\ref{Table 1} provides the details of the two processed datasets.

\noindent \textbf{Baseline}: We compare DCLMDB with existing causal debiasing methods. Causal methods are often used as supplementary techniques alongside backbone recommendation models. In the experiments, all baseline methods use two commonly used backbone recommendation models, namely MF and LightGCN. We will compare our approach against the following six causal methods:
    \begin{itemize}
		\item[\textbullet] \textbf{IPS}~\cite{schnabel2016recommendations}: IPS addresses popularity bias in models by tackling the imbalance in the long-tail distribution within observational data. Specifically, it assigns the inverse of an item's popularity as its weight, elevating the significance of less popular items and reducing the weight of more popular ones.
       \item[\textbullet] \textbf{IPS-C}~\cite{bottou2013counterfactual}: This method imposes a cap on the maximum value of IPS weights to reduce the variance in the overall weight distribution.
		\item[\textbullet] \textbf{IPS-CN}~\cite{gruson2019offline}: IPS-CN applies normalisation to reduce variance in weight distribution.
		\item[\textbullet] \textbf{CauseE}~\cite{bonner2018causal}: CauseE constructs a small unbiased dataset, and employs matrix factorisation on both datasets to derive two sets of embeddings, then applies $L_1$ or $L_2$ regularisation to align these embeddings, enforcing their similarity.
		\item[\textbullet] \textbf{DICE}~\cite{zheng2021disentangling}: DICE utilises Structural Causal Modeling (SCM)~\cite{pearl2009causality} to model user-item interactions as an interplay between ``interest'' and ``conformity''. It constructs specific training samples based on collision effects, thereby disentangling the embedding of each user and item into separate ``interest" and ``conformity" components to eliminate conformity bias.
        \item[\textbullet] \textbf{DCCL}~\cite{zhao2023disentangled}: This study critiques DICE's use of specific data to disentangle ``interest'' and ``conformity'', suggesting it may introduce noise. DCCL adopts contrastive learning to address data sparsity and disentangle ``interest'' and ``conformity''.
    \end{itemize}

We evaluate the Top-K recommendation performance on implicit feedback, a commonly used setting in recommender systems. Top-K refers to the K items that the recommender system deems most relevant or attractive to the user, where K represents the number of items in the recommendation list. We utilize three frequently used evaluation metrics: \textit{Recall}, \textit{Hit Rate} (\textit{HR}), and \textit{NDCG}. In the experiments, the reported results represent the optimal performance achieved by each method under its parameter settings. Additionally, we counted the degree of improvement of Recall for each method compared to the backbone, denoted by ``Imp.''.

\subsection{Comparison of Experimental Results}
We report the results of DCLMDB and all baseline approaches in Tables \ref{Table 2} and \ref{Table 3} on the two real-world datasets.

	% \usepackage{multirow}
    \begin{table*}[htbp]
		\centering	
		\begin{tabular}{cc|cccccccc}
			\hline
			\multicolumn{2}{c|}{\multirow{2}{*}{Dataset}} &
			\multicolumn{8}{c}{Netflix} \\ \cline{3-10} 
			\multicolumn{2}{c|}{} &
			\multicolumn{4}{c|}{Top-K=20} &
			\multicolumn{4}{c}{Top-K=50} \\ \hline
			\multicolumn{1}{c|}{Backbone} &
			Method &
			\multicolumn{1}{c}{Recall$\uparrow$} &
			\multicolumn{1}{c}{HR$\uparrow$} &
			\multicolumn{1}{c}{NDCG$\uparrow$} &
			\multicolumn{1}{c|}{Imp.$\uparrow$} &
			\multicolumn{1}{c}{Recall$\uparrow$} &
			\multicolumn{1}{c}{HR$\uparrow$} &
			\multicolumn{1}{c}{NDCG$\uparrow$} &
			Imp.$\uparrow$ \\ \hline
			\multicolumn{1}{c|}{\multirow{8}{*}{MF}} &
			Original &
			\multicolumn{1}{c}{0.1154} &
			\multicolumn{1}{c}{0.5262} &
			\multicolumn{1}{c}{0.0959} &
			\multicolumn{1}{c|}{--} &
			\multicolumn{1}{c}{0.1947} &
			\multicolumn{1}{c}{0.6804} &
			\multicolumn{1}{c}{0.1200} &
			-- \\ %\cline{2-10} 
			\multicolumn{1}{c|}{} &
			IPS &
			\multicolumn{1}{c}{0.1043} &
			\multicolumn{1}{c}{0.4816} &
			\multicolumn{1}{c}{0.0831} &
			\multicolumn{1}{c|}{-9.61\%} &
			\multicolumn{1}{c}{0.1843} &
			\multicolumn{1}{c}{0.6541} &
			\multicolumn{1}{c}{0.1081} &
			-5.34\%\\ %\cline{2-10} 
			\multicolumn{1}{c|}{} &
			IPS-C &
			\multicolumn{1}{c}{0.1109} &
			\multicolumn{1}{c}{0.5012} &
			\multicolumn{1}{c}{0.0885} &
			\multicolumn{1}{c|}{-3.89\%} &
			\multicolumn{1}{c}{0.1901} &
			\multicolumn{1}{c}{0.6649} &
			\multicolumn{1}{c}{0.1133} &
			-2.36\%\\ %\cline{2-10} 
			\multicolumn{1}{c|}{} &
			IPS-CN &
			\multicolumn{1}{c}{0.1179} &
			\multicolumn{1}{c}{0.5298} &
			\multicolumn{1}{c}{0.1032} &
			\multicolumn{1}{c|}{+2.17\%} &
			\multicolumn{1}{c}{0.2040} &
			\multicolumn{1}{c}{0.6817} &
			\multicolumn{1}{c}{0.1290} &
			+4.78\%\\ %\cline{2-10}  
			\multicolumn{1}{c|}{} &
			CausE &
			\multicolumn{1}{c}{0.0996} &
			\multicolumn{1}{c}{0.4832} &
			\multicolumn{1}{c}{0.0838} &
			\multicolumn{1}{c|}{-13.69\%} &
			\multicolumn{1}{c}{0.1751} &
			\multicolumn{1}{c}{0.6460} &
			\multicolumn{1}{c}{0.1057} &
			-10.07\%\\ %\cline{2-10} 
			\multicolumn{1}{c|}{} &
			DICE &
			\multicolumn{1}{c}{\underline{0.1265}} &
			\multicolumn{1}{c}{\underline{0.5535}} &
			\multicolumn{1}{c}{\underline{0.1069}} &
			\multicolumn{1}{c|}{\underline{+9.62\%}} &
			\multicolumn{1}{c}{\underline{0.2166}} &
			\multicolumn{1}{c}{\underline{0.7065}} &
			\multicolumn{1}{c}{\underline{0.1341}} &
			\underline{+11.24\%} \\ %\cline{2-10}
            \multicolumn{1}{c|}{} &
			DCCL &
			\multicolumn{1}{c}{0.1231} &
			\multicolumn{1}{c}{0.5433} &
			\multicolumn{1}{c}{0.1023} &
			\multicolumn{1}{c|}{+6.67\%} &
			\multicolumn{1}{c}{0.2091} &
			\multicolumn{1}{c}{0.6955} &
			\multicolumn{1}{c}{0.1283} &
			+7.40\%\\ %\cline{2-10}
			\multicolumn{1}{c|}{} &
			 DCLMDB &
			\multicolumn{1}{c}{\pmb{0.1313}} &
			\multicolumn{1}{c}{\pmb{0.5731}} &
			\multicolumn{1}{c}{\pmb{0.1126}} &
			\multicolumn{1}{c|}{\pmb{+13.78\%}} &
			\multicolumn{1}{c}{\pmb{0.2204}} &
			\multicolumn{1}{c}{\pmb{0.7152}} &
			\multicolumn{1}{c}{\pmb{0.1391}} &
			\pmb{+13.20\%} \\ \hline
			\multicolumn{1}{c|}{\multirow{8}{*}{LightGCN}} &
			Original &
			\multicolumn{1}{c}{0.1149} &
			\multicolumn{1}{c}{0.5272} &
			\multicolumn{1}{c}{0.0962} &
			\multicolumn{1}{c|}{--} &
			\multicolumn{1}{c}{0.2009} &
			\multicolumn{1}{c}{0.6718} &
			\multicolumn{1}{c}{0.1219} &
			--\\ %\cline{2-10} 
			\multicolumn{1}{c|}{} &
			IPS &
			\multicolumn{1}{c}{0.1139} &
			\multicolumn{1}{c}{0.5211} &
			\multicolumn{1}{c}{0.0970} &
			\multicolumn{1}{c|}{-0.87\%} &
			\multicolumn{1}{c}{0.1953} &
			\multicolumn{1}{c}{0.6718} &
			\multicolumn{1}{c}{0.1219} &
			-2.79\%\\ %\cline{2-10} 
			\multicolumn{1}{c|}{} &
			IPS-C &
			\multicolumn{1}{c}{0.1164} &
			\multicolumn{1}{c}{0.5256} &
			\multicolumn{1}{c}{0.0997} &
			\multicolumn{1}{c|}{+1.31\%} &
			\multicolumn{1}{c}{0.1985} &
			\multicolumn{1}{c}{0.6777} &
			\multicolumn{1}{c}{0.1249} &
			-1.19\%\\ %\cline{2-10} 
			\multicolumn{1}{c|}{} &
			IPS-CN &
			\multicolumn{1}{c}{0.0776} &
			\multicolumn{1}{c}{0.4160} &
			\multicolumn{1}{c}{0.0691} &
			\multicolumn{1}{c|}{-32.46\%} &
			\multicolumn{1}{c}{0.1549} &
			\multicolumn{1}{c}{0.6009} &
			\multicolumn{1}{c}{0.0922} &
			-22.90\%\\ %\cline{2-10} 
			\multicolumn{1}{c|}{} &
			CausE &
			\multicolumn{1}{c}{0.0919} &
			\multicolumn{1}{c}{0.4594} &
			\multicolumn{1}{c}{0.0751} &
			\multicolumn{1}{c|}{-20.02\%} &
			\multicolumn{1}{c}{0.1690} &
			\multicolumn{1}{c}{0.6336} &
			\multicolumn{1}{c}{0.0987} &
			-15.88\%\\ %\cline{2-10} 
			\multicolumn{1}{c|}{} &
			DICE &
			\multicolumn{1}{c}{\underline{0.1410}} &
			\multicolumn{1}{c}{\underline{0.5848}} &
			\multicolumn{1}{c}{\underline{0.1212}} &
			\multicolumn{1}{c|}{\underline{+22.72\%}} &
			\multicolumn{1}{c}{\underline{0.2343}} &
			\multicolumn{1}{c}{\underline{0.7290}} &
			\multicolumn{1}{c}{\underline{0.1490}} &
			\underline{+16.63\%} \\ %\cline{2-10} 
            \multicolumn{1}{c|}{} &
			DCCL &
			\multicolumn{1}{c}{0.1200} &
			\multicolumn{1}{c}{0.5389} &
			\multicolumn{1}{c}{0.1012} &
			\multicolumn{1}{c|}{+4.44\%} &
			\multicolumn{1}{c}{0.2027} &
			\multicolumn{1}{c}{0.6902} &
			\multicolumn{1}{c}{0.1261} &
			+0.90\%\\ %\cline{2-10}
			\multicolumn{1}{c|}{} &
			DCLMDB &
			\multicolumn{1}{c}{\pmb{0.1442}} &
			\multicolumn{1}{c}{\pmb{0.5948}} &
			\multicolumn{1}{c}{\pmb{0.1250}} &
			\multicolumn{1}{c|}{\pmb{+25.50\%}} &
			\multicolumn{1}{c}{\pmb{0.2383}} &
			\multicolumn{1}{c}{\pmb{0.7325}} &
			\multicolumn{1}{c}{\pmb{0.1526}} &
			\pmb{+18.62\%}\\ \hline
		\end{tabular}
        \caption{The performance of all methods on Netflix. DCLMDB’s improvement on Netflix is highly significant, with a $p$-value of 0.0007 in pairwise comparison to the 12 results of the second-best method.}
        \label{Table 3}
    \end{table*}
	
From Tables~\ref{Table 2} and \ref{Table 3}, CausE does not demonstrate any performance improvement, likely, because it only coarsely aligns the biased embeddings with the unbiased ones and fails to eliminate the impact of confounders. DCCL, which applies contrastive learning to DICE, yields even poorer results. This is due to the fact that DCCL's starting point is to address the problem of data sparsity, so it directly employs the original training data.  
Although both DCCL and DCLMDB apply contrastive learning, DCLMDB performs debiasing on user-item pairs rather than using data augmentation on the user or item itself. In contrast, our DCLMDB framework outperforms all the baselines, demonstrating significant improvements across all metrics on both datasets. For example, when using MF as the backbone on the Movielens-10M dataset, DCLMDB shows an increase of over 35\% in the Recall@20 metric and more than 26\% improvement in Recall@50. Similarly, DCLMDB achieves notable enhancements across all three metrics on the Netflix dataset. It is worth noting that the comparison of DCLMDB with the second-best method, DICE, on the two datasets yields $p$-values of 0.004 and 0.007, respectively, indicating a statistically significant improvement.

DCLMDB is a versatile framework that can be seamlessly integrated into various recommendation models, regardless of the backbone.  It consistently delivers optimal results across different datasets and excels in three metrics. According to the results presented in Tables \ref{Table 2} and \ref{Table 3}, methods that do not involve the embedding layer, such as IPS, show inconsistent performance. While IPS improves upon the MF model in the Movielens-10M, it negatively impacts the performance of the MF model on the Netflix. This inconsistency is also observed in other non-embedding layer baselines, likely due to their overdependence on dataset distribution. In contrast, DCLMDB, which applies debiasing and disentangling at the embedding layer and leverages causal graph analysis, consistently enhances recommendation performance across various datasets and backbone models.

\subsection{Ablation Experiments}
\begin{figure}[t]
	\centering
	\includegraphics[width=0.96\columnwidth]{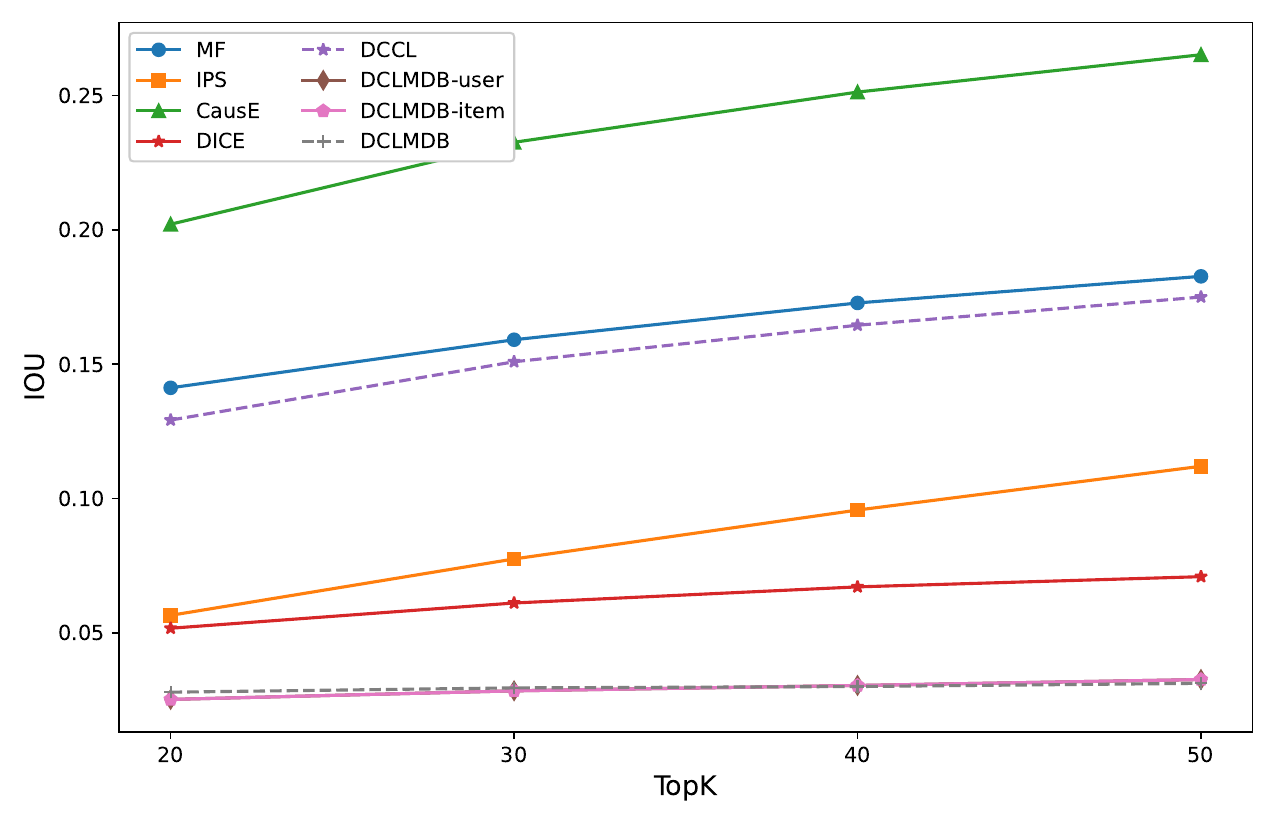}
	\caption{Overlapped items with popular items. A higher IOU indicates that the recommendation result is more similar to the recommended top popular items.}
	\label{Fig7}
\end{figure}

 In the ablation experiments, we introduced two variants of DCLMDB: DCLMDB-user and DCLMDB-item. These variants only eliminate conformity bias and popularity bias during training, respectively. We performed ablation experiments on DCLMDB to determine the validity of the debiasing effect of each component. We evaluated all methods, including  DCLMDB, its variant versions, as well as the baseline methods, using the Intersection Over Union (IOU)~\cite{zheng2021disentangling}, an evaluation metric that measures the degree of overlap between their recommended items and popular items. The results on the Movielens-10M dataset, as shown in Fig.~\ref{Fig7}, indicate that both DCLMDB-user and DCLMDB-item exhibit lower overlap ratios between its recommended items and popular items compared to other baselines. This outcome highlights their effectiveness in removing both popularity bias and conformity bias, thereby affirming the efficacy of our approach. Further analysis of the curves in Fig.~\ref{Fig7} reveals that the curve for our method remains relatively flat, whereas the curves for other baselines show a marked upward trend. This observation suggests a significant diminution in the debiasing effect of the baseline algorithms as the Top-K value increases. In contrast, our method maintains a consistent debiasing effect across various Top-K values, demonstrating its robustness. In conclusion, DCLMDB not only excels in eradicating popularity bias and conformity bias but also maintains stable debiasing effects under varying Top-K conditions.
\begin{figure}[t]
    \centering
    \includegraphics[width=\columnwidth]{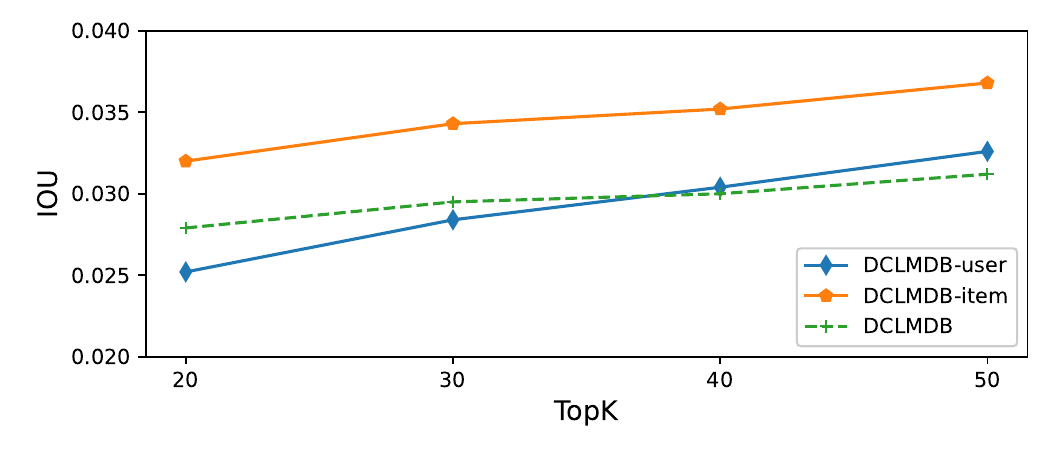}
    \caption{The IOU ratios of DCLMDB and its variants. }
    \label{Fig8}
\end{figure}

Fig.~\ref{Fig8} illustrates the IOU ratios of DCLMDB and its variant methods, showing more clearly the debiasing effect between them. In Fig.~\ref{Fig8}, it is remarkable to notice that DCLMDB-user and DCLMDB-item show different degrees of weakening of their debiasing effects when the number of recommended items increases. However, the debiasing effect of the DCLMDB method remains stable without significant fluctuations. It shows that the DCLMDB method is robust. This further demonstrates that the strategy of simultaneous debiasing on both the user side and the item side is effective and can make the debiasing effect more stable. Ablation experiments for the Netflix dataset are provided in the Appendix.

\section{Conclusion}
In this paper, we first presented an analysis of the impact of popularity and conformity on recommender systems from a causal perspective using the proposed causal graph. Guided by causal-based analysis and diverging from existing studies on eliminating popularity bias and conformity bias, we formulated the click prediction model using do-calculus. To address both popularity bias and conformity bias simultaneously, we proposed DCLMDB, a novel debiased contrastive learning method for recommender systems. DCLMDB utilises initial item and user embeddings as negative samples to derive two sets of embeddings specifically designed to eliminate popularity and conformity biases, and it accomplishes this by implementing targeted interventions in the model's training process, using contrastive learning. %The experiments conducted on two real-world datasets have demonstrated the robustness of the proposed causal solution and the effectiveness of DCLMDB. 
The experimental results conducted on two real-world datasets indicate that our proposed DCLMDB is more effective and consistent in eliminating popularity bias in both items and users compared to other benchmark models. This leads to superior recommendation performance, as evidenced by our findings.

\bibliography{aaai25}

\clearpage

\appendix

\section{Preliminaries}
In this section, we will introduce the fundamental concepts of causal inference related to our main manuscript.

Causal graphs use Directed Acyclic Graphs (DAGs) to present causal relationships between variables, where nodes represent variables and edges represent relationships between them. Specifically, there are three classical DAGs to describe causal relationships between variables: chain $A\rightarrow B \rightarrow C$, fork  $A\leftarrow B \rightarrow C$, and collider $A\rightarrow B \leftarrow C$. In the causal DAG $A\rightarrow B \rightarrow C$, $A$ influencing $C$ through intermediary $B$. In the causal DAG $A\leftarrow B \rightarrow C$, $B$ is referred to as the confounder or common cause of $A$ and $C$, i.e., $B$ influences both $A$ and $C$, resulting in an correlation $A$ and $C$. Note that there is not imply a direct causal relationship between $A$ and $C$.  $A\rightarrow B \leftarrow C$ is a collider structure, where $A$ and $C$ are independent of each other but jointly influence the collision node $B$. $A$ and $C$ exhibit correlation when conditioned on $B$. 

Do-calculus is a derivation system consisting of three derivation rules. Before introducing the rules, we illustrate two special subgraphs. In a causal DAG $\mathcal{G}$ with three arbitrarily disjoint sets of nodes $X$, $Y$, and $Z$, we denote by $\mathcal{G}_{\overline{X}}$ the graph obtained by deleting from $\mathcal{G}$ all arrows pointing to nodes in $X$. Likewise, we denote by $\mathcal{G}_{\underline{X}}$ the graph obtained by deleting from $\mathcal{G}$ all arrows emerging from nodes in $X$.

\begin{theorem}[Rules of do-calculus~\cite{pearl2009causality}]
Let $\mathcal{G}$ be DGA associated with a causal model, and let $P(\cdot)$ stand for the probability distribution induced by that model. For any disjoint subsets of variables X, Y, Z, and W, we have the following rules. 

\begin{enumerate}
    \item Insertion/deletion of observations: 
    \begin{align}
    \label{eq002}
    &P(y\mid do(x),z,w)  \notag \\
    &= P(y\mid do(x),w)\quad if(Y\CI Z\mid X,W)_{\mathcal{G}_{\overline{X}}}.
    \end{align}
    
    \item Action/observation exchange:
    \begin{align}
    \label{eq003}
    &P(y\mid do(x),do(z),w)  \notag \\
    &= P(y\mid do(x),z,w)\quad if(Y\CI Z\mid X,W)_{\mathcal{G}_{\overline{X}\underline{Z}}}.
    \end{align}
    
    \item Insertion/deletion of actions:
    \begin{align}
    \label{eq004}
    &P(y\mid do(x),do(z),w)  \notag \\
    &= P(y\mid do(x),w)\quad if(Y\CI Z\mid X,W)_{\mathcal{G}_{\overline{X},\overline{Z(W)}}},    
    \end{align}
    where Z(W) is the set of Z-nodes that are not ancestors of any W-node in $\mathcal{G}_{\overline{X}}$.
\end{enumerate}

\end{theorem}

For example, we perform $do(A=a)$ on the causal DAG $A\leftarrow B \rightarrow C$, which denotes the intervention of setting the variable $A$ to be $a$ and cutting off the path $B \rightarrow A$ on the causal DAG.

\section{Experiments}

\subsection{Experimental Settings}
\noindent \textbf{Parameter Settings}: To ensure fair comparisons, we standardised the parameter counts across all methods. For models utilising DICE~\cite{zheng2021disentangling} and DCCL~\cite{zhao2023disentangled}, we set the embedding size to 64, since they comprise two concatenated sets of embeddings. For the other models, we maintained a consistent embedding size of 128. In the DCLMDB method, the hyperparameters $\alpha$ and $\beta$ were set to 0.05 and 0.005, respectively, for the MF-based model, and to 0.5 and 0.005 for the GCN-based model. We employed the Adam optimiser for updating model weights, with an initial learning rate of 0.001 and a batch size of 128. All models use BPR~\cite{rendle2012bpr} loss as the objective function for click prediction. All models were executed on an NVIDIA A100 (40GB RAM) GPU. To assess model performance and validate the effectiveness of our approach, we utilised three widely recognised metrics in the recommendation systems field: \textit{Recall}, \textit{Hit Ratio} (\textit{HR}), and \textit{NDCG}.

\subsection{Correlation Analysis between Popularity and Conformity Biases}

In this section, we conduct experiments to analyze and demonstrate the correlation between popularity and conformity biases, thereby validating the rationality of the causal graph proposed in our experiments.

\begin{figure}[htbp]
        \begin{center}
		\subfigure[]{
            \includegraphics[width=0.8\columnwidth]{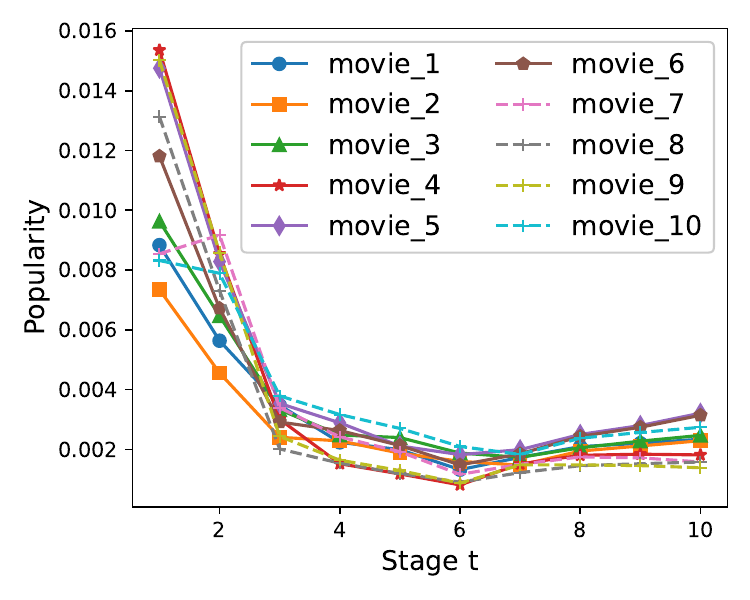}
            }
            \subfigure[]{
            \includegraphics[width=0.8\columnwidth]{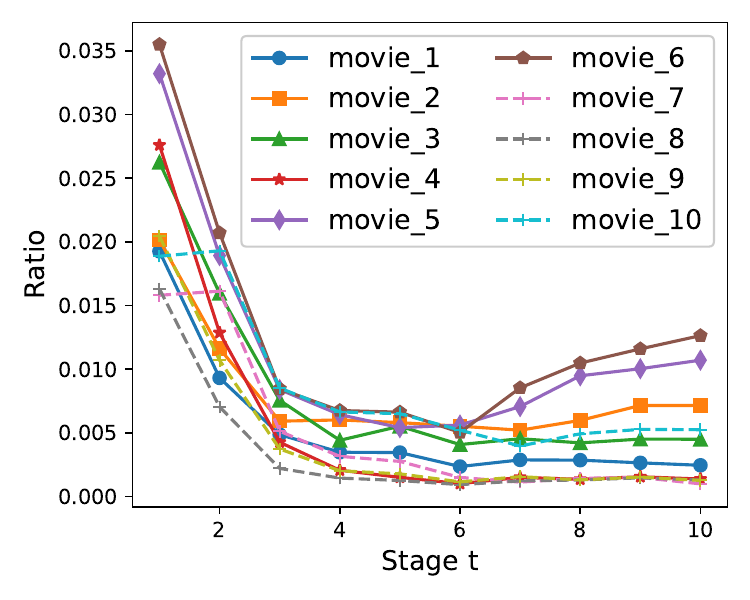}
            }
		\caption{Correlation analysis between popularity bias and conformity bias. (a) Local praise rate change curve at each stage; (b) Local popularity change curve at each stage.}
		\label{Fig5}           
        \end{center}
	\end{figure}

Let $D$ represent the historical observational data between items and users, $R$ denote the count of five-star ratings for each item, and $I$ represent the set of all items. We divided the Movielens-10M dataset into $T$ stages in timestamp order. For each stage $t$, we computed the local positive rating rate and local popularity, defined as follows:
	\begin{equation}
		\begin{array}{l}
			r_{i}^{t}= R_i^t / \displaystyle\sum_{j\in I}R_j^t,
		\end{array}
	\end{equation}
	\begin{equation}
		\begin{array}{l}
			m_{i}^{t}= D_i^t / \displaystyle\sum_{j\in I}D_j^t,
		\end{array}
	\end{equation}
\noindent where $R_i^t$ denotes the count of five-star ratings received by item $i$ at stage $t$, while $D_i^t$ represents the total number of interactions with item $i$ at the same stage. The terms $r_{i}^{t}$ and $m_{i}^{t}$ correspond to the conformity and popularity of item $i$ at stage $t$, respectively. To illustrate these concepts, we randomly selected ten items with relatively high interaction counts and compared the trends of their local positive rating rates and local popularity, as shown in Fig. \ref{Fig5} (a) and (b). The trends in the local positive rating rates and popularity among these items displayed a notable similarity.

 \begin{figure}[htbp]
        \centering
        \includegraphics[width=\columnwidth]{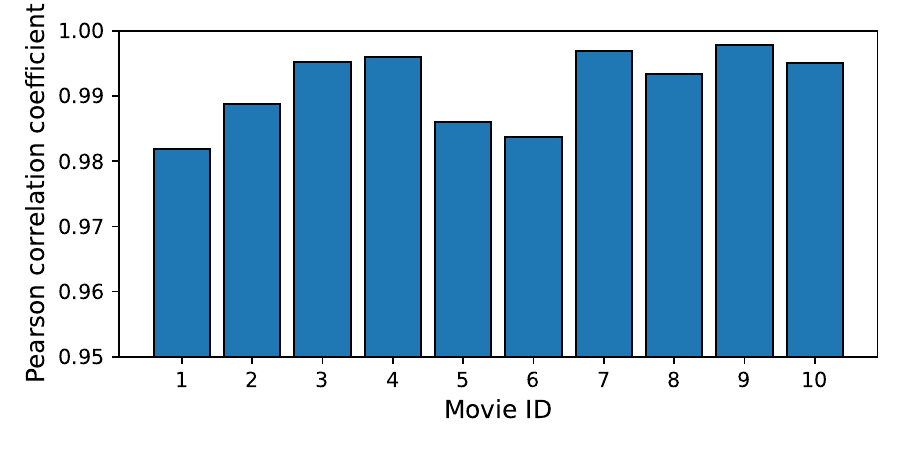}
        \caption{The Pearson correlation coefficient between the local praise rate and popularity.}
        \label{Fig6}
    \end{figure}
    
To scientifically demonstrate their correlation, we calculated the Pearson correlation coefficient for these ten items, as shown in Fig. \ref{Fig6}. This coefficient measures the strength of the relationship between two variables and ranges from $-1$ to $1$, where values closer to $1$ indicate a stronger positive correlation. The Pearson correlation coefficients, as displayed in Fig. \ref{Fig6} for the trends of local positive rating rates and local popularity among these ten items, all exceed $0.98$. These results suggest an exceptionally strong correlation between popularity bias and conformity bias. They imply that these biases often coexist and should be addressed simultaneously rather than focusing on only one aspect of the bias.

\subsection{Performance under Intervened Training Data}
\begin{figure}[!h]
    \centering
    \includegraphics[width=\columnwidth]{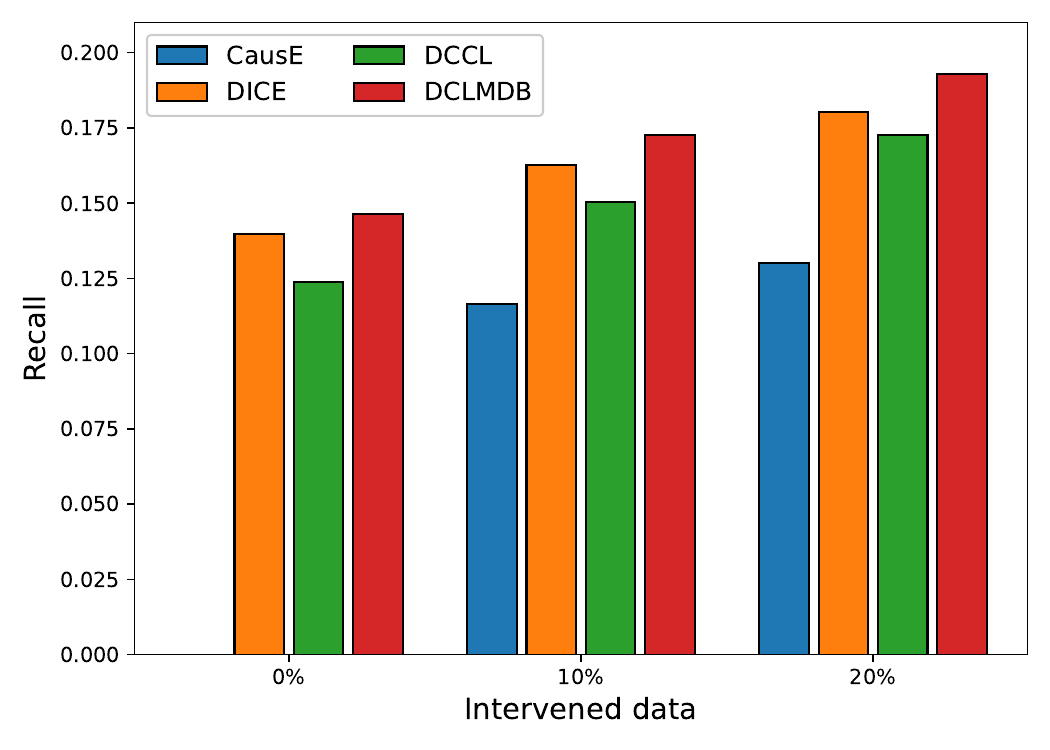}
    \caption{Performance comparison between different algorithms under different proportions of intervened training data.}
    \label{Fig9}
\end{figure}
In this section, we assess the performance of DCLMDB across various proportions of intervention data. In the prior experiment, due to CausE's stringent requirements, we were compelled to merge biased data (60\%) with intervention data (10\%) to form a training set. As the training data includes intervention data, DCLMDB's effectiveness might be enhanced by it. Additionally, in actual recommendation systems, acquiring intervention data often involves randomly suggesting items to users, an expensive practice that can degrade user experience and increase user attrition. Thus, to further validate DCLMDB's effectiveness and superiority, we utilised 0\%, 10\%, and 20\% intervention data proportions in the Movielens-10M training set, employing MF as the backbone. Fig.~\ref{Fig9} illustrates the performance of DCLMDB, DCCL, DICE, and CausE under varying intervention data proportions. Notably, when the intervention data proportion is 0\%, CausE is unable to yield results as it necessitates intervention data in its training set. The figure distinctly shows that DCLMDB markedly surpasses the other baselines both in the absence and presence of various scales of intervention data, robustly confirming its effectiveness and superiority.

\subsection{Ablation Experiments of the Netflix Dataset}
\begin{figure}[htbp]
	\centering
	\includegraphics[width=\columnwidth]{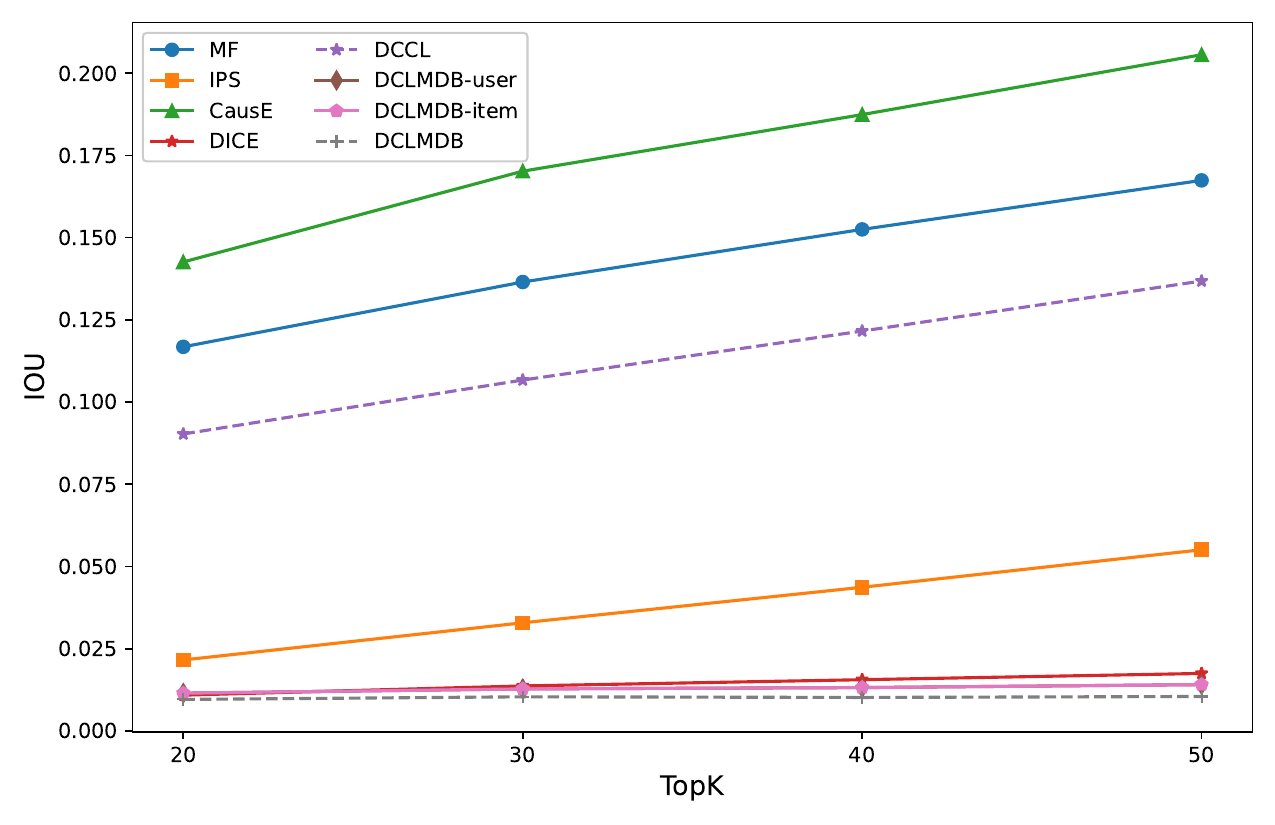}
	\caption{Overlapped items with popular items on the Netfilx dataset. A higher IOU indicates that the recommendation result is more similar to the recommended top popular items.}
	\label{DCLMDB_iou_on_nf}
\end{figure}

We will perform ablation experiments on the Netflix dataset. As with the experimental results on the Movielens-10M dataset, the overlap between recommended and popular entries is lower for DCLMDB-user and DCLMDB-item compared to other baselines. This result illustrates their effectiveness in eliminating prevalence bias and conformity bias, further validating the effectiveness of our method. Meanwhile, as can be seen in Fig~\ref{DCLMDB_iou_on_nf}, the IOU curves of DCLMDB and its variant methods are relatively flat, which illustrates the robustness of the debiasing ability of our method.
\begin{figure}[htbp]
    \centering
    \includegraphics[width=\columnwidth]{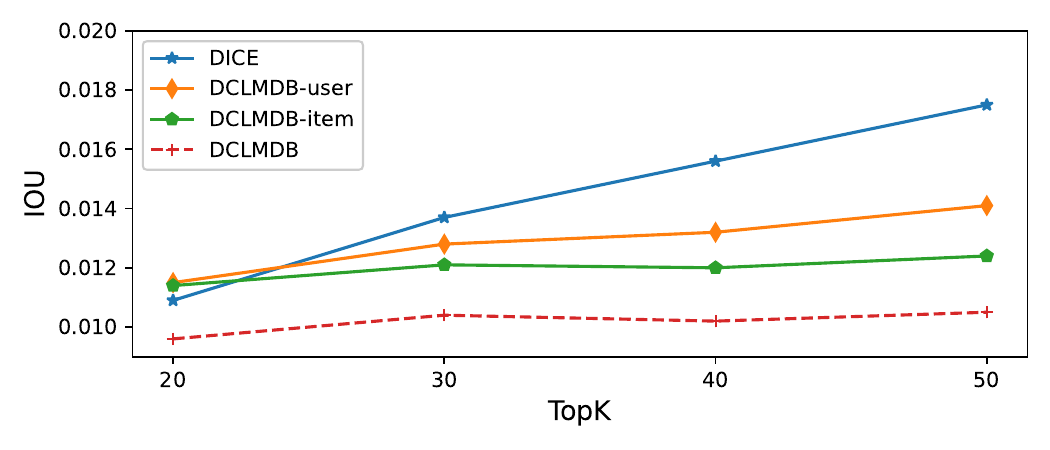}
    \caption{The IOU ratios of DICE, DCLMDB and its variants.}
    \label{DCLMDB_local_iou_nf}
\end{figure}

Fig.~\ref{DCLMDB_local_iou_nf} shows the IOU ratios of DICE, DCLMDB, and its variants to more clearly demonstrate the debiasing effect between them. In Fig.~\ref{DCLMDB_local_iou_nf}, DCLMDB-user, DCLMDB-item, and DICE all show different decreases in their debiasing power compared to DCLMDB as the number of recommended items increases. However, the debiasing effect of the DCLMDB method remains stable without significant fluctuations. This phenomenon is consistent with the results of our ablation experiments on the Movielens-10M dataset, which further verifies the strong robustness of the DCLMDB method. It also demonstrates again that the strategy of simultaneous debiasing on the user side and the item side is effective, which can make the debiasing effect more stable.
\end{document}